# Swarm Intelligent Algorithm For Re-entrant Hybrid Flow shop Scheduling Problems

**Zhonghua Han**

Faculty of Information and Control Engineering,

Shenyang Jianzhu University,

Shenyang, Liaoning, China

Department of Digital Factory,

Shenyang Institute of Automation,

Chinese Academy of Sciences,

Shenyang, Liaoning, China

Email: xiaozhonghua1977@163.com

**Xutian Tian\***

Faculty of Information and Control Engineering,

Shenyang Jianzhu University,

Shenyang, China

Email: tytxt5992@foxmail.com

\*Corresponding author

**Xiaoting Dong**

Faculty of Electrical Engineering,

Sichuan College of Architectural Technology,

Deyang, Sichuan ,China

Email: dxt199211@163.com

**Fanyi Xie**

Faculty of Information and Control Engineering,

Shenyang Jianzhu University,

Shenyang, China

Email: xfy895623@163.com

Biographical notes:

Zhonghua Han received his PhD degree at Shenyang Institute of Automation, China in 2014. He is currently a Professor with the Faculty of Information and Control Engineering, Shenyang Jianzhu University, Shenyang, China. His main research includes production and operation management, integrated technology of automation system in enterprise, and the engineering application research of production scheduling method.

XuTian Tian is a master candidate from the Faculty of Information and Control Engineering, Shenyang Jianzhu University, Shenyang, China. He currently does his study under the supervising of Zhonghua Han.His main research area is production scheduling.

Xiaoting Dong is a postgraduate student from the Faculty of Information and Control Engineering, Shenyang Jianzhu University, Shenyang, China. Her main research area is production scheduling.

Fanyi Xie is a master candidate from the Faculty of Information and Control Engineering, Shenyang Jianzhu University, Shenyang, China. She currently does her study under the supervising of Zhonghua Han.Her main research area is production scheduling.

**Abstract:** In order to solve Re-entrant Hybrid Flowshop (RHFS) scheduling problems and establish simulations and processing models, this paper uses Wolf Pack Algorithm (WPA) as global optimization. For local assignment, it takes minimum remaining time rule. Scouting behaviors of wolf are changed in former optimization by means of levy flight, extending searching ranges and increasing rapidity of convergence. When it comes to local extremum of WPA, dynamic regenerating individuals with high similarity adds diversity. Hanming distance is used to judge individual similarity for increased quality of individuals, enhanced search performance of the algorithm in solution space and promoted evolutionary vitality.A painting workshop in a bus manufacture enterprise owns typical features of re-entrant hybrid flowshop. Regarding it as the algorithm applied target, this paper focus on resolving this problem with LDWPA (Dynamic wolf pack algorithm based on Levy Flight). Results show that LDWPA can solve re-entrant hybrid flowshop scheduling problems effectively.

## 1 Introduction

Kumar(1993)first proposed the RHFS line(Li and Wang, 2010) as the third type of scheduling problem that distinguishes it from Flowshop and Jobshop. Reentrant manufacturing system have all jobs of reentrant working procedure featuring re-entering the waiting matching area. It leads to instability scheduling problems of reentrant workshop and obstruct, making scheduling problems of reentrant manufacturing workshop more complicated than general issue. Many tasks, various processes, multi work stations, and different processing time of jobs in the same working procedure are features of hybrid flowshop problems (HFSP) (Han et al., 2016; Karimi et al., 2011). If reentrant manufacturing lines appears in the RHFS, more complex production process, sharply increased production loads, added instability and unbalanced equipment loading and other questions greatly result in HFS difficulty. RHFS scheduling problem (Ying et al., 2015)is a typical kind of NP-hard problem. RHFS scheduling problems can be found in semi-conduct manufacturing, bus production and steel smelting. Multi chip packages of semiconductor packaging production line, multiple chip bonding line and multi-sticking crape masking process of a painting workshop have obvious characters of RHFS.

Many scholars have made varying degrees of progress in research on RHFS scheduling. Liu et al.(2011) and Zhu and Chen(2018) use particle swarm with fast convergence and strong global searching ability of genetic algorithm for an exchanged result, designing a genetic particle swarm optimization for reentrant manufacturing dispatch algorithm. (Fan et al., 2012) study aero-engine reassembly line in a workshop after multi decomposition. With the stochastic matrix coding method, crossover method and variation method, an optimization scheduling method based on genetic algorithm is proposed to find the optimal method of assembly workshop scheduling. (Lin et al., 2016) aimed to meeting the characteristics of the dynamic reach ability and re-entry of jobs in mold heat treatment, weighted tardiness and energy consumption index are established as optimization objectives, having tempering process in mould heat treatment been in rolling schedule for heuristic algorithm. As for obstacles of two classes of RFFS, (Sangsawang et al., 2015) investigates GA based on fuzzy logic controller and PSO about Cauchy distribution to solve the problem.

According to relevant references in recent years, existing works(Collart and Verschueren, 2014;Lin D et al., 2012) have supplied optimization algorithm to Re-entrant Flowshop scheduling problems and achieve good results. Algorithm is more accurate for optimization after improvement algorithm and combined with various algorithms, especially the traditional meta heuristic algorithm and improved algorithm which includes (Re-entrant flexible Flowshop, (RFFS) scheduling problems, there are few references about RHFS scheduling problems. This paper investigates LDWPA (The Dynamic wolf pack algorithm based on Levy Flight), and show LDWPA has faster speed and higher accuracy compared with other existing algorithm

## 2 Describe and mold about RHFS dispatch problems

As for RHFS scheduling problems, it can be described as jobs processed in working procedures. Jobs are produced according to the specified sequence of operations in the process flow. One working procedure of all contains at least one work station. There are differences about process time of jobs. At least one process with re-entrant features work station must be supplied to job's process flow .After jobs traversing all process are re-inserted into the queue to be processed. Jobs with the least remaining processing time are chosen in free work station. Scheduling problems are about work station distribution, process, start time and completion time. Process must follow the rules below:

1   Every workstation can manufacture only one job at a time.
2   Every procedure of one job are processed in one workstation.
3   The whole procedure of job process are not allow to be interrupted.
4   Transit time of jobs is neglected during processes.

During the scheduling process of RHFS, in the number $nrm$ non-reentrant manufacturing process, each workstation consists of a set of the same work stations $M_1 \sim M_{nrm}$. In the number $rm$ re-entrant manufacturing phase, each workstation composes a set of the same work stations $M_{nrm+1} \sim M_{nrm+rm}$; the job $J_i$ needs to re-enter this phase $rts_i$ times, while different job has its own re-entrant times. In this problem, each job in accordance with the specified process flow goes through various workstations for processing. Each job will be processed for $nrm + rm \times rts_i$ times. Flow process chart is shown in Figure 1.

**Figure 1** the chart of process sequence

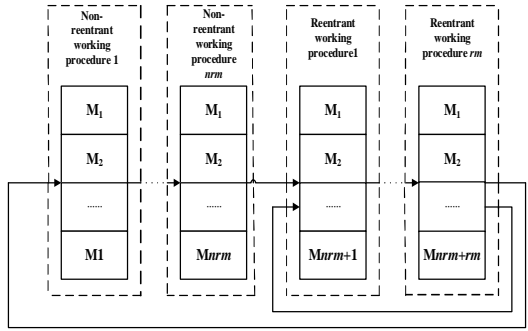

## 2.1 Mathematic simulations and processing model description

### 2.1.1 Parameter setting

In this section, there are many mathematical variables of the re-entrant hybrid flow shop for establishing its mathematics planning model**s.**

$n$ means the largest number of processing jobs

$J_i$ means the $i$ th job $i \in \{1,......,n\}$.

$m$ means the whole number of working procedures.

$OP_j$ means $j$ process.

$M_j$ means the largest work station of $OP_j$ working procedure.

$WS_{j,k}$ means the $k$ th work station of $OP_j$ working procedure.

$FL_q$ means the process flow of jobs, it is processed working procedure collection of jobs according to the process flow.

$rm$ means the re-entrant processes number in production line.

$nrm$ means the non-re-entrant processes numbers in production line.

$rts_i$ means $J_i$ job experience times of reentrant working procedure.

$om_i$ means the whole number of processed jobs in production procedure $FL_q$. Cumulative counting about the whole number of the re-entrant working procedure, $om_i \geq m$.

$l$ means the sequence number of being processed working procedure in $FL_q$ $l \in \{1,......,om_i\}$ process flow.

$S_{i,j,k}^l$ means $J_i$ job, the $OP_j$ working procedure in the process flow of $FL_q$, the start process time.

$C_{i,j,k}^l$ means $J_i$ job, the $OP_j$ working procedure in the process flow of $FL_q$, the finished process time.

$WT_{i,j,k}^l$ $J_i$ job, the $OP_j$ working procedure in the process flow of $FL_q$, the process time.

$WS_{j,k}$ means the serial number of job in the $k$ th station of $OP_j$ working procedure.

### 2.1.2 Assumed variables and basic constrained relationships

In this section, many mathematical formulas are established with the variable set up in the previous section, through which constraint relations in the process of scheduling optimization are included, reflecting the procedure and features of scheduling optimization in the Re-entrant Hybrid flow shop.

$$At_{i,j,k}^{t_i} = \begin{cases} 0 & \text{Workpiece } J_i \text{ doesn't process on the work station } WS_{j,k} \text{ of } OP_j \text{ during the number } t_i \text{ working procedure.} \\ 1 & \text{Workpiece } J_i \text{ process on the work station } WS_{j,k} \text{ of } OP_j \text{ during the number } t_i \text{ working procedure} \end{cases}$$

The variable $WS_{j,k}$ represents if the jobs are in the $FL_q$ process. In the $l$ th processed of $OP_j$ working procedure is about the process situation of $WS_{j,k}$ job.

$$C_{i,j,k}^l = S_{i,j,k}^l + WT_{i,j,k}^l,$$
$$i \in \{1,......,n\}, l \in \{1,......,om_i\}, j \in \{1,......,m\} \quad (1)$$

$$C_{i,j,k}^l \geq S_{i,j,k}^l,$$
$$i \in \{1,......,n\}, l \in \{1,......,om_i\}, j \in \{1,......,m\} \quad (2)$$

$$\sum_{t_i=1}^{om_i} At_{i,j,k}^{t_i} = om_i,$$
$$i \in \{1,......,n\}, l = om_i \quad (3)$$

$$n_{j,k} = \sum_{i=1}^{n} \sum_{t_i=1}^{om_i} At_{i,j,k}^{t_i} \quad (4)$$

$$om_i = nrm + rm \times rts_i \quad (5)$$

Formula (1) indicates relationships among job's start process time, process time and finished process time in its processing flow with the re-entrant process. Formula (2) indicates limited relationships of the same job between continuous process start and finished process time. Formula (3) indicates every job should experience their whole process flow. Formula (4) indicates that it is necessary to accumulate number of repetitively processed jobs when counting the number of jobs processed in a workstation in re-entrant workshop.（5）indicates that the whole number of jobs in $FL_q$ equals to the whole number of $J_i$ in non-reentrant and reentrant working procedures.

## 3 WPA designs

The Dynamic wolf pack algorithm based on Levy Flight (LDWPA) after utilizing WPA(Yang et al., 2007;Zhou et al., 2013) is in this paper, which includes two improvements.

1. Scouting behaviors based on Levy Flight: scouting behaviors are about position changes from present situation to new one. The search ranges of this kind of local random walking is focused in such small scope. Basically, the Levy Flight is a random walk combines of long step and small step. The Levy Flight has a larger search ranges and ability compared with random walking because of directed long step. Founding wolf position is improved according to levy flight, the wolf rushed to its own searching area of low-probability with a long step in solution space. It also extends searching range for a development of wolf optimization.

2. The species dynamic mechanism based on hanming distance: After a certain number of iterations, WPA is shows the phenomenon of evolutionary stagnation. The reason is that with the increased iterations, more similar and even the same genes of individuals and useless communication between individuals attribute to this result. Therefore, a dynamic updating population method based on stagnation evolution iterations is introduced. If the optimum value of stagnated revolution iterations exceeds threshold, the hanming distance can be used to judge the similarity among individuals and abandon the individuals with the highest similarity of best one and left new individual with low similarity and large differences, making a diversity, highlighting local extremum and keeping vitality of algorithm.

## 3.1 Population dynamic renewal mechanism

The basic concept and characteristics of Hanming distance are charted in 3.1.1 firstly. In 3.1.2 is about real examples of LDWPA to reflect objectively the distance of long code word based on Hanming distance. In order to judge differences among individuals in population on the condition that the iterative process falls to hysteresis, the method of using Hanming distance to judge the difference among individuals for less difference one is introduced in 3.1.3. Besides, it can be used to increase new individuals with more differences, enhance diversity during evolution and maintain evolutionary vitality.

### 3.1.1 Hanming distance

Hanming distance (Harada et al., 2017; Atallah and Duket 2011; Haider et al., 2015) is a basic concept in information theory, describing the distance of two long code words.

$$D(x, y) = \sum_{k=1}^{n} x_k \oplus y_k,$$

$$x = (x_1, x_2, ..., x_n), \quad y = (y_1, y_2, ..., y_n) \tag{6}$$

$\oplus$ represents XOR operation, $x_k \in \{0,1\}$, $y_k \in \{0,1\}$, $D(x, y)$ means the whole number of two code words in the same position. It reflects the differences between two code words and evidences for the similarity.

### 3.1.2 The individual similar judgment based on hanming distance

In WPA, the individual is the sequence of real numbers and matrix. The individual similarity can be found from the individual distribution of population. In view of the fact that similarity judgment method is in the evolve algorithm, in order to reduce the complexity of the operation and improve the time efficiency. Through the similarity $D$ between wolves, that is, the ratio of the same gene in the two individuals to the total number of genes in the individual. Then compare with the threshold to find similar individuals. Firstly, individual is ranked by fitness. The similarity $D$ between the leader wolf and other individuals in wolf pack can be calculated. If $D$ is greater than the threshold $Rt$, two individuals are similar individuals which leads to a temporary subpopulation $StPop1$. $StPop2$ is built in the rest individuals by repeating the above operations for better individuals till the end. In every temporary subpopulation, many similar individuals can be ranked according to adjustment degree. Partial individuals can be reserved based on $Kr$ ratio of better individual as its revolution adaptation out of others.

$$d_{i,t_i} = \begin{cases} 0 & \lfloor a_{i,t_i} \rfloor \neq \lfloor a'_{i,t_i} \rfloor \\ 1 & \lfloor a_{i,t_i} \rfloor = \lfloor a'_{i,t_i} \rfloor \end{cases}$$

The variate $d_{i,t_i}$ means upper line of two different individuals $X_{np}$ and $X_{np}'$. The responding gene section $\lfloor a_{i,t_i} \rfloor$ and $\lfloor a'_{i,t_i} \rfloor$ reflects situation of the upper line. If the situation is same, $d_{i,t_i} = 1$; if not, $d_{i,t_i} = 0$.

$$N_D = \sum_{i=1}^{n} \sum_{t_i=1}^{om_i} D_{i,t_i} \tag{7}$$

$N_D$ in (7) means the number of two individual with the same gene.

$$D = \frac{N_D}{N_{Gene}} \tag{8}$$

$D$ is the similarity of individual and the proportion between $N_D$ of two individuals with

the same gene section and the whole number of $N_{Gene}$. If it surpasses the threshold $Rt$, two individuals is similar.

### 3.1.3 Dynamic renewable similar individual

The individual with a high similarity can be found after every generation when it comes to a certain iterations. Reserving superb individuals, eliminating similar individuals, then generate new individuals with large differences from the population and replace individuals with high similarity in order to preserve good distribution of population in solution space and strengthen the global searching ability. The purpose of using starting generation $StartGen$ mainly consider that the population still has good evolutionary vitality at the early stage of evolution. When it comes to a certain number of iterations, the vitality will decrease and then the operation of dynamic updating population is introduced which reduce computation of the algorithm during the entire evolution process.

### 3.2 Scouting behaviors based on levy flight

This section combines the dynamic renewal populations based on Hanming distance mentioned above and the Scouting behavior depended on Levy flight. By creating the constraint relation of different variables and listing the changing variables in the evolution process. This part describes all steps of the LDWPA proposed in this paper in detail and makes the logical relations more clearly based on the flow chart of the algorithm.

French mathematician Levy put forward a probability distribution called levy distribution in 1930s, which leads to a large amount of scholars' researches. Till now, scholars can prove that foods searching routines of many animals and insects (albatross, bee and fruit fly) are the same with levy distribution. It also explains many natural random phenomenon, such as Brownian movement. Levy flight(Palyulin et al., 2014)conforms to random searching routines of the levy distribution. It is a walking way combined with short distance searching and occasionally long distance walking. Levy fight is adopt to upgrade many population optimization. Many scholars (Ibrahim, 2016; Ehsan et al., 2013) utilize it in improvement of informational interaction among individuals and searching for the optimal solution in the solution space. In a conclusion, there is a possibility of population individual advancing to former small searching range, expanding the range. It has achieved satisfactory results in increasing population diversity and also improved the rapidity and veracity of algorithm. Therefore, the use of Swarm Intelligent Algorithm based on Levy flight makes it easier to jump out of the local extremum which can effectively enhance the algorithm's optimization ability.

The new method of competitive wolf's position based on levy flight scouting behaviors.

$$x_i^{(t+1)} = x_i^{(t)} + stepa \times levy(u,v) \qquad (9)$$

$i = 1......n$

$x_i^{(t)}$ means the $t$ th position of competitive wolf; $\oplus$ is multiplication; $step\Gamma$ is step element.

Levy flight basically is a random step which conforms to levy distribution. It has been achieved because of the complicated levy distribution, which leads to the frequent usage of Mantegna algorithm. Mantegna algorithm about levy flight is as followed.

$$levy(\{\}) \sim u = \frac{\mathsf{w}u}{|v|^{\frac{1}{\mathsf{s}}}} \qquad (10)$$

$$\dagger_u = \left\{ \frac{\Gamma(1+\mathsf{s})\sin(\frac{f\mathsf{s}}{2})}{\Gamma[1+\mathsf{s}]\mathsf{s} \cdot 2^{\frac{(1-\mathsf{s})}{2}}} \right\}^{\frac{1}{\mathsf{s}}}, u \sim N(0,\dagger_u^2) \qquad (11)$$

$$\dagger_v = 1, v \sim N(\dagger_v{}^2) \qquad (12)$$
$$s = 1.5$$

A larger searching range and improvement of WPA can been achieved by advancing to a small possibility range of levy flight.

### 3.3 LDWPA operations

Step1 The parameter of original algorithm and $Np$ wolves form the initial population, initializing positions of every wolf. The maximum number of iteration is $Genmax$, the number of scouting wolf is $q$, each scouting wolf hunt prey in $h$ directions, the maximum searches is $SCmax$ and $(SCmax \leq 15)$. The searching step $stepa$ and move step are $stepb$. There are $bw$ wolves will be eliminated in each iteration.

Step2 $q$ wolves are used for the leader wolf by hunting. The $cwp_k$ wolf's scouting behaviors searches for better positions based on levy flight and (9).

Step3 Other wolves which fail to implement scouting behaviors followed the best wolf as the leader wolf. It's position is changed through $x_{kd}' = x_{kd} + rand \bullet stepb \bullet (x_{ld} - x_{kd})$: rand is a random number between (0,1). $x_{ld}$ is $d$ dimensional vector of the leader wolf.

Step4 Regenerating population according to the wolf distribution principle, removing the worst $bw$ wolf and creating new $bw$ wolves by initializing position through
$$X_{bw_k} = x_{min} + rand(x_{max} - x_{min})$$

$x_{max}$ and $x_{min}$ represents the upper limitation and lower limitation of value. The leader wolf randomly moves to find the food, notifies the other wolves around the prey by howling, and the other wolves surround the leader wolf. For this behavior, a random number $r_k$ generated in [0,1] is generated. If $r_k$ is smaller than $_n$ (a preset threshold), the wolf $x_k$ does not move. If $r_k$ is greater than $_n$. The wolf $x_k$ surrounds the prey with the leader wolf.

The new position of $k$ wolf is
$$x_k' = x_k + rand \times ra$$
$$ra(t) = ra_{min} \times (x_{max} - x_{min}) \times \exp(\frac{\ln(ra_{min}/ra_{max}) \times t}{maxt})$$

Step5 The best individual and revolution limitation can be judged or not. If it meets the condition, it will be finish; if it does not meet the condition, it will continue.

Step6 If the population revolution is none of newer best individual, it is necessary to stop evolve iteration. $stopgen = stopgen + 1$.

Step7 The evolvement reaches $StartGen$ iteration at the beginning of new population similar individual or not. If $gen < StartGen$, which indicates that it does not reach the calculating of iteration and turn to Step 2; Otherwise, it starts operational formula of dynamic update similar individuals.

Step8 Calculating the similarity $SI$ among individuals. Individuals that $SI$ are greater than the threshold $Rt$ are made into a temporary subpopulation, Subpopulation of these temporary subpopulation is $ltPop$. Each member $Nt$ in $ltPop$ contains several subpopulations $stPop$. The number of each temporary subpopulation $stPop$ is $stnp$.

Step9 Assuming the individual count variable of a temporary new population is $inp$. According to better individual reserve percentage $Kr$. Preserving better individuals of each temporary subpopulation are $Kr \times stnp$. That is, preserving individuals with fitness values that match evolutionary trends. The best individual is put into a new group $newPop$ and $inp = Kr \times stnp$.

Step10 If the individual number of $newPop$ reaches to $Np$ and set a new individual count variable $tnp = 0$.

Step11 If the new individual number $tnp$ bigger or the same with $(Np - inp)$ and then to the step 13; if not, go on to the Step 12.

Step12 New individuals came into being in this way to judge the similarity between the new individual and individual of $newPop$. This new individual will be abandoned as long as the similarity between the new individual and individual of $newPop$ above the threshold $Rt$. This new individual will be reserved as long as the similarity between the new individual and individual of $newPop$ lower or the same with the threshold $Rt$. This individual is added to $newPop$, $tnp = tnp + 1$, and go on to Step 11.

Step13 Regarding $Pop$ as a new group Pop with $gen = gen + 1$ is the first step before Step 2.

$q$ is 5, searching direction $h$ is 4, searching step $stepa$ is $0.6 \times |x_{max} - x_{min}|$ and moving step $stepb$ is 0.3. $t$ is the current iteration number, $max_t$ is the maximum number of iterations, and $ra_{max}$ =400 is the maximum surrounding steps. $ra_{min}$ =0.5 is minimum surrounding steps. Similarity threshold $Rt$ is 0.6. Better individual reserve ratio $Kr$ is 0.3.

**Figure 2** The chart of algorithm sequence

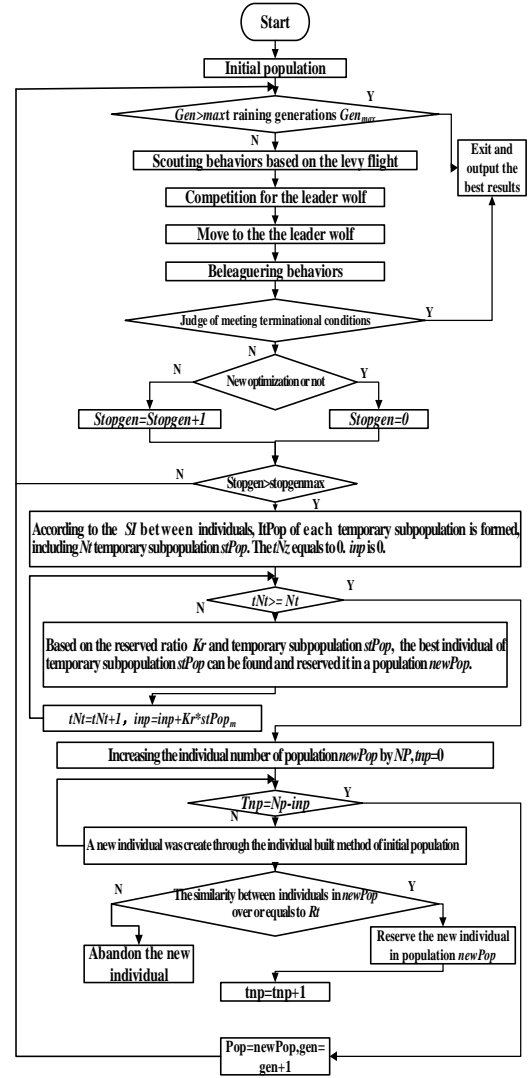

## 4 Simulation experiments

In this section, The LDWPA is used to optimize benchmarks of different scales in this section. The optimization results are compared with these obtained of GA, WPA and WOA simulation examples in order to prove the effectiveness and superiority of the LDWPA algorithm.

**4.1 Algorithm analysis**

To verify effectiveness of LDWPA algorithm mentioned in the former part, the flexible flow shop examples are used to compare performance of the algorithm under the real facts that researches on re-entrant hybrid flow shop scheduling optimization problem is still in the primary stage, and lack of examples. Adopt the benchmark published by (Carlier and Néron, 2000), this section includes three hard instances: j15c5d2, j15c5d3, j15c5d4 and three easy instances: j15c5a4, j15c5a5, j15c5b1.

Comparing the LDWPA algorithm with the NEH algorithm mentioned in (Ribas et al., 2010) research and the WPA algorithm, the best lower bound $LB'$ is known as the best optimal scheduling result. The effectiveness of the algorithm is evaluated by the following indexes: deviation $d$ refers to the deviation between $C_{max}$ and $LB'$. Among in $d$ refers to:

$$d = ((C_{max} - LB')/LB') \times 100\%$$

Based on CPU 2.5GHz and 4G internal memory, MATLAB environment is the prerequisite of GA, WOA, WPA, LDWPA algorithms in this paper.

In table 1, it can be concluded from the table that three algorithms have obtained best optimization results in the easy instances. However, the advantages of LDWPA algorithm is more obvious in hard problems. The optimization effects of WPA and NEH are nearly the same as for the easy instances. In the difficult instances, such as J15c5d3, the effects of WPA optimization algorithm are worse than NEH because that WPA algorithm is easy to the local extremum and lack of evolutionary vitality. Both the optimal solutions and average optimal solutions from LDWPA are better than those from the contrast algorithm NEH, which shows that the LDWPA algorithm mentioned in this paper greatly improves the overall optimization effects of the WPA algorithm, effectiveness and superiority of LDWPA algorithm.

In this paper, two small scale examples and five large scale examples mentioned in (Sun et al., 2017) research are used for simulation test. By using GA algorithm mentioned in Sun Y's paper, the optimization results of different scale examples are compared with those obtained by WOA, WPA and LDWPA selected in this paper. This proves the validity and superiority of the algorithm in dealing with various scale data.

**Table 1**     the results of instance in three algorithm

| Num | BH | LB' | NEH $C_{max}$ | NEH $C'_{max}$ | NEH d | WPA $C_{max}$ | WPA $C'_{max}$ | WPA d | LDWPA $C_{max}$ | LDWPA $C'_{max}$ | LDWPA d |
|---|---|---|---|---|---|---|---|---|---|---|---|
| 1 | d2 | 82 | 92 | 92.4 | 12.1 | 91 | 92.5 | 10.9 | 84 | 85.4 | 3.6 |
| 2 | d3 | 77 | 88 | 88.3 | 14.2 | 89 | 89,6 | 15.5 | 83 | 84.2 | 7.7 |
| 3 | d4 | 61 | 89 | 90.2 | 45.9 | 88 | 88.9 | 44.2 | 84 | 85.1 | 37.7 |
| 4 | a5 | 130 | 131 | 130.4 | 0 | 130 | 130 | 0 | 130 | 130 | 0 |
| 5 | a4 | 156 | 156 | 156.1 | 0 | 156 | 156 | 0 | 156 | 156 | 0 |
| 6 | b1 | 170 | 170 | 170 | 0 | 170 | 170 | 0 | 170 | 170 | 0 |

**Figure 3**   The Iterative Curve Diagram of Four kinds of algorithms to solve the Scheduling Problems of instance of J15c5d3

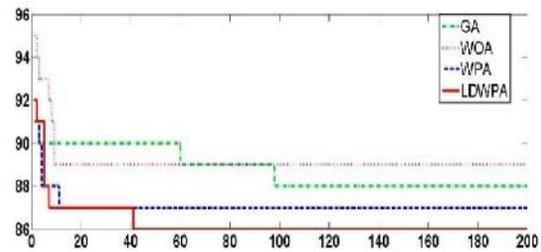

Figure 3 is a proper revolution curve graph from optimization algorithm data of GA, WOA, WPA and LDWPA. With the development of exercise in figure 3, the fitness value of four groups is smoothly small till the stable situation. Optimized curve convergence rate is slow in GA algorithm. The optimization is inferior to other three methods in limited evolution frequency. The optimized curve convergence rate with WOA algorithm is rapidly and stop within 20[th] generation, leading to the local extremum and bad evolution situation. The bad fitness reflects the fast convergence rate、local extremum and bad optimization. The optimized WPA algorithm between 10[th] generation and sharp decreased 20[th] generation is superior to whales optimization algorithm in optimization speed. It also exists the weak local extremum capacity and fragile vitality. The worse first generation optimization fitness through the LDWPA algorithm is the same with the initial generation among 20[th] generation. Thus, the optimized curve by WPA is nearly reach the fitness. The optimization speed is faster than WOA and WPA featuring fast speed. It is concluded that the Levy flight optimizes the local searching and fosters the algorithm optimization. With the dynamic new population, algorithm among 10[th] generation and 40[th] generation keeps evolution after its local extremum and jump from the local extremum for a better fitness value 86. The dynamic upgraded population greatly improves WPA of being in local extremum.

**Figure 4**   The chart of four type algorithms

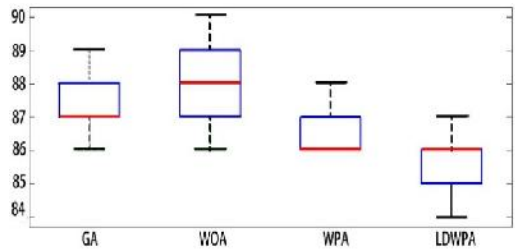

Figure 4 is the Box-plot about fitness value after 20 times same date by means of optimizing GA, WOA, WPA, LDWPA. The abscissa is the algorithm of example date and the ordinate is about fitness values. The box's tendency is down in the process of optimization from the Box-plot above, testifying the whole result is up in the comparable four algorithms. The minimum of improved algorithm is smaller than other three algorithms, resulting to the higher possibility of qualitative results in the improved algorithm. 1[st] quartile and maximum are equal to 3[rd] quartile of WPA and minimum, which is also smaller than the range of median of GA、3[rd] quartile and fitness value of algorithm. It can be concluded that the overall results' quality through the LDWPA algorithm is superior to other two algorithms. The unusual number 90 is related to fitness value of initial population. The worst result appears when the fitness value of initial population is 20.

From the partial solution examples in table 2, the fitness value of every algorithm is similar for the small range examples. The LDWPA is still better to get results, whose average is also less than that of other algorithms. That is to say, the LDWPA has good optimization ability in small range of date. It is necessary to notice that the former WPA , with the feature of being into local extremum, exists phenomena of lacking evolve vitality and relying on optimal value of initial population. From the chart, GA algorithm has a feature of jumping out of local extremum. But results of GA algorithm is better than those of WPA with enough generations. With population dynamic mechanism and the levy flight in neighbor searching, the LDWPA's results is superior to GA algorithm in medium and big date, which proves the efficiency and stability of this algorithm in the paper.

**Table 2**   The results of examples in different algorithms

| | | LDWPA | | WPA | | WOA | | GA | |
|---|---|---|---|---|---|---|---|---|---|
| | BH | Best fitness | Average fitness | Best fitness | Average fitness | Best fitness | Average fitness | Best fitness | Average fitness |
| 1 | J15c5d2 | 84 | 85.2 | 87 | 87.7 | 88 | 88.9 | 88 | 89.5 |
| 2 | J15c5d3 | 82 | 82.4 | 84 | 85.1 | 87 | 88.1 | 90 | 91.1 |
| 3 | j80c4a1 | 1415 | 1433.2 | 1723 | 1735.2 | 1713 | 1754.4 | 1699 | 1710.1 |
| 4 | j80c8a2 | 1776 | 1794.7 | 2033 | 2056.4 | 2040 | 2061.1 | 2033 | 2056.9 |
| 5 | j120c4a1 | 2077 | 2094.6 | 2692 | 2711.2 | 1616 | 1625.3 | 2688 | 2697.3 |
| 6 | J120c4a2 | 2218 | 2236.9 | 2670 | 2689.3 | 2677 | 2687.5 | 2654 | 2672.7 |
| 7 | j120c8a1 | 2477 | 2494.4 | 2840 | 2845.7 | 2877 | 2897.4 | 2826 | 2830.4 |

### 4.2 Experimental analysis on LDWPA results and RHFS dispatching problems

This section uses the LDWPA algorithm, which has proved its superiority in the former section to optimize the operation section of multiple masking procedures in the painting shop with the characteristics of re-entrant hybrid flow shop in the actual bus manufacturing enterprises. By establishing different evaluation indexes, the superiority of LDWPA algorithm to solve the re-entrant hybrid flow shop scheduling optimization problem can be proved.

**Figure 5** Deployment Diagram of Work Station in Multi-Sticking Crape Masking Procedure

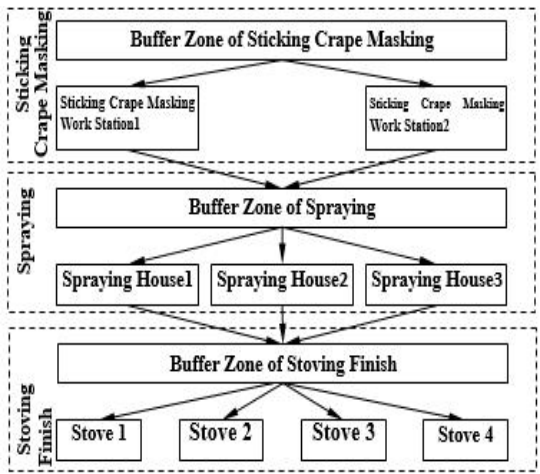

In the working multi-sticking crape masking working procedures of spraying workshop in automobile manufacturing factory. The technological persons disperse color ribbon into different colors according to the difficulty of color ribbon models and spray various color ribbon patterns on bus external parts. Spraying one kind of color after sticking color ribbon(crape masking)、spraying and baking paint and the whole pattern come into being after several times of spraying. Therefore, multi-sticking crape masking working procedure is a typical re-entrant stage, including many working procedures and work stations. This working procedure has obvious features of hybrid flowshop because that various abilities of technological persons attribute to different process times. Aimed at scheduling problems of multi-sticking crape masking working procedures in spraying workshop, this paper optimizes it with LDWPA and resolve scheduling problems in RHFS.

Some automobile manufacturing enterprises provide scheduling date about the practical process information of multi-sticking crape masking working procedures in coating workshops. This figure 5 is about the distribution of multi-sticking crape masking working procedure.

The result is about the scheduling problems of multi-sticking crape masking working procedures in 15 automobiles' bodies. The working procedures of $\{OP_1, OP_2, OP_3\}$ mean stickingcolor ribbon、spraying and baking paint in multi-sticking crape masking working procedures. These three working procedures contain parallel work stations of 2,3,4. painting housefor spraying as well as painting house for baking paint are treated as work stations in thescheduling procedure.

$\{J_1, J_2, J_3, J_4, J_5, J_6, J_7, J_8\ J_9, J_{10}, J_{11}, J_{12}, J_{13}, J_{14}, J_{15}\}$ represent jobs of the fifteen types of car body to be processed. Fifteen cars belong to fifteen types manufactured by its order. X in the list means automobile process time of multi-sticking crape masking working procedures. The manufacture time in all working procedure is different because of different types of buses. There is no people in the process of baking paint in painting house. Therefore, the baking paint time in different painting house about the same type of automobiles is no different and working time difference is related to types.

**Table 3** Time of multi-sticking crape masking working procedures（min）

| Working Procedure | | Work station | Time of multi-sticking crape masking working procedure | | | | | | | | | | | | | | |
|---|---|---|---|---|---|---|---|---|---|---|---|---|---|---|---|---|---|
| | | | $J_1$ | $J_2$ | $J_3$ | $J_4$ | $J_5$ | $J_6$ | $J_7$ | $J_8$ | $J_9$ | $J_{10}$ | $J_{11}$ | $J_{12}$ | $J_{13}$ | $J_{14}$ | $J_{15}$ |
| | | | $FL_1$ | $FL_2$ | $FL_3$ | $FL_4$ | $FL_5$ | $FL_6$ | $FL_7$ | $FL_8$ | $FL_9$ | $FL_{10}$ | $FL_{11}$ | $FL_{12}$ | $FL_{13}$ | $FL_{14}$ | $FL_{15}$ |
| 1stC OLO RRI BBO N | $OP_1^1$ | $WS_{1,1}$ | 12 | 15 | 15 | 12 | 15 | 15 | 15 | 20 | 12 | 18 | 15 | 15 | 15 | 20 | 15 |
| | | $WS_{1,2}$ | 15 | 18 | 12 | 15 | 10 | 12 | 20 | 18 | 15 | 15 | 12 | 18 | 20 | 18 | 18 |
| | $OP_2^2$ | $WS_{2,1}$ | 15 | 18 | 20 | 18 | 15 | 18 | 20 | 20 | 15 | 18 | 15 | 18 | 20 | 25 | 25 |
| | | $WS_{2,2}$ | 18 | 20 | 16 | 20 | 15 | 15 | 20 | 20 | 20 | 22 | 15 | 15 | 25 | 25 | 18 |
| | | $WS_{2,3}$ | 15 | 15 | 15 | 18 | 18 | 20 | 22 | 22 | 18 | 20 | 18 | 20 | 20 | 20 | 20 |
| | $OP_3^3$ | $WS_{3,1}$ | 20 | 20 | 20 | 20 | 20 | 20 | 25 | 25 | 25 | 25 | 20 | 20 | 25 | 25 | 25 |
| | | $WS_{3,3}$ | | | | | | | | | | | | | | | |
| 2nd COL ORR IBB ON | $OP_1^4$ | $WS_{1,1}$ | 15 | 15 | 12 | 10 | 12 | 15 | 15 | 15 | | | | | | | |
| | | $WS_{1,2}$ | 12 | 12 | 15 | 12 | 10 | 15 | 10 | 12 | | | | | | | |
| | $OP_2^5$ | $WS_{2,1}$ | 18 | 15 | 15 | 15 | 15 | 20 | 15 | 15 | | | | | | | |
| | | $WS_{2,2}$ | 15 | 18 | 18 | 15 | 10 | 15 | 15 | 15 | | | | | | | |
| | | $WS_{2,3}$ | 18 | 15 | 15 | 15 | 15 | 15 | 20 | 20 | | | | | | | |
| | $OP_3^6$ | $WS_{3,1}$ | 20 | 18 | 18 | 18 | 18 | 20 | 20 | 20 | | | | | | | |
| | | $WS_{3,3}$ | | | | | | | | | | | | | | | |
| 3rdC OLO RRI BBO N | $OP_1^7$ | $WS_{1,1}$ | 10 | 12 | 12 | | | | | | | | | | | | |
| | | $WS_{1,2}$ | 12 | 10 | 10 | | | | | | | | | | | | |
| | $OP_2^8$ | $WS_{2,1}$ | 10 | 15 | 12 | | | | | | | | | | | | |
| | | $WS_{2,2}$ | 15 | 10 | 15 | | | | | | | | | | | | |
| | | $WS_{2,3}$ | 15 | 15 | 10 | | | | | | | | | | | | |
| | $OP_3^9$ | $WS_{3,1}$ | 18 | 18 | 18 | | | | | | | | | | | | |
| | | $WS_{3,3}$ | | | | | | | | | | | | | | | |

The GA, WPA and LDWPA algorithms are used to compare the different evaluation indexes. The makespan $C_{max}$ is used as the fitness value function of the overall optimization algorithm during the optimization process. At the same time, many evaluation indexes related to the practical application of the production line are established. These include the total load balance cost $TLB$, the total workstation free time $TWT$ and the total equipment utilization rate $FUR$. Except for the total equipment utilization $FUR$, the other evaluation index value is smaller and better. The evaluation indicators are described as following:

Total load balance $TLB$.

$$\overline{WT_j} = \left( \frac{\sum_{j=1}^{M_j} \sum_{i=1}^{n} \sum_{l=1}^{om} \left( WT_{i,j,k}^l \cdot At_{i,j,k}^{t_i} \right)}{M_j} \right) \quad (13)$$

$$TLB = \sum_{j=1}^{m} \sqrt{\sum_{k=1}^{M_j} \left( \left( \sum_{l=1}^{om} \left( WT_{i,j,k}^l \cdot At_{i,j,k}^{t_i} \right) - \overline{WT_j} \right)^2 \right)} \quad (14)$$

$\overline{WT_j}$ in formula (13) means the average processing time of $M_j$ work stations of $Op_j$ working procedure. By summing up the total processing time at each station $WS_{j,k}$ of the procedure $Op_j$ and the average processing time $\overline{WT_j}$ of the work stations and the square of the average processing time $\overline{WT_j}$. The load balance cost $TLB_j$ is established in formula (14).

The total load balance cost $TLB$ is the sum of the cost of the load balance of all working procedures during the whole process.

Total equipment utilization ratio $FUR$

$$FUR = \frac{\sum_{k=1}^{M_j}\left(\sum_{j=1}^{m}\left(\sum_{i=1}^{n}\left(\sum_{l=1}^{om}WT_{i,j,k}^{l}\right)\right)\right)}{\sum_{j=1}^{m}\left(\sum_{k=1}^{M_j}\left(\max\left\{C_{i,j,k}^{l}\cdot At_{i,j,k}^{ti}\right\}-\min\left\{S_{i,j,k}^{l}\cdot At_{i,j,k}^{ti}\right\}\right)\right)} \quad (15)$$

$FUR$ in formula (15) is the total equipment utilization ratio of all stations in the flexible flow shop. It is also the ratio of the all effective processing time of work stations and the span of the work station working time. This time span is a period from the beginning of the first processing task to the final processing task.

**Table 4**   The results of evaluation indexes in three algorithms

| evaluation index | GA | | WPA | | LDWPA | |
|---|---|---|---|---|---|---|
| | Best fitness | Average fitness | Best fitness | Average fitness | Best fitness | Average fitness |
| $C_{max}$ | 255 | 257.6 | 262 | 264.75 | 247 | 247.8 |
| $TWT$ | 488 | 490.1 | 503 | 505.2 | 481 | 482.4 |
| $FUR$ | 73.92% | 73.45% | 72.93% | 72.56% | 75.01% | 74.85% |
| $TLB$ | 187.60 | 189.27 | 193.15 | 197.33 | 184.33 | 186.15 |

Optimization performances of each optimization algorithm is analyzed from the data in Table 4. The average of maximum completion time $C_{max}$ from LDWPA algorithm is better than that of GA and WPA algorithm. The three algorithms are similar to each other in terms of furniture utilization ratio $FUR$, but the optimization effects of LDWPA is still slightly higher than that of the other two algorithms. The optimization effect of WPA algorithm for total load balance $TLB$ and free time of total work station $TWT$ is obviously worse than that of GA algorithm. However, the optimization effect of LDWPA algorithm on every evaluation index is better than that of GA algorithm, which shows that the optimization ability and the quality of solution obtained by LDWPA algorithm are greatly improved compared with WPA algorithm. LDWPA can optimize every index, which shows that LDWPA can reduce the free time of work station better, arrange the processing task more reasonably, make the equipment load more balanced, and have better optimization performance.

Figure 6 shows the gantt of the re-entrant hybrid flow shop based on the actual data during production process. The work $J_7$, traversing the process route, passes through the work station $\{WS_{12}, WS_{21}, WS_{32}, WS_{12}, WS_{21}, WS_{33}\}$, and the work $J_2$ is produced in station $WS_{11}$ when it firstly enters the Re-entrant working procedure $OP_1$. The job is produced in the work station $WS_{12}$ when it first comes to re-entrant working procedure $OP_1$. Besides, the two stations in the working procedure $OP_1$ is different. Both of them indicates that the scheduling of work production tasks is constantly changing during the ongoing production procedure, reflecting that Re-entrant Hybrid flow shop scheduling problem is a dynamic problem. Special attention is paid to the work $J_2$ and the work $J_7$ entering firstly the second working procedure $OP_2$ of the re-entrant working procedure part at the same time. According to the local assignment rule specified in this paper, the work with the smaller remaining has a higher priority to enter the station for processing. According to table 2, the remaining of $J_7$ is shorter than that of $J_2$, so the work $J_7$ is first processed by the free station $WS_{21}$.

**Figure 6** Gantt results of Re-entrant Hybrid Flowshop

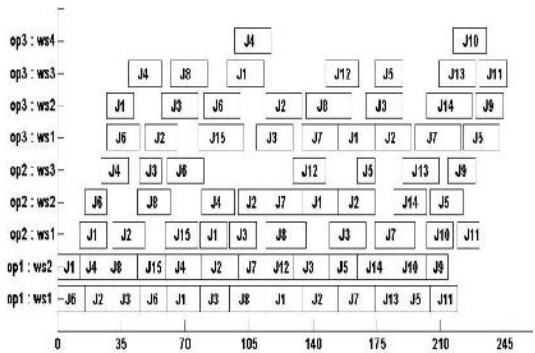

## 5 Conclusion

Based on multi-sticking crape masking in spraying workshop of automobile manufacture enterprises, this paper investigates the RHFS searching problems. In view of reentrant process system in practical society, this paper also raises RHFS dispatch searching problem with reentrant system and describes this problem. Mathematics models come into being with the target of minimizing the complete time. This paper highlights the LDWPA algorithm. Improving the scouting behaviors of WPA with the Levy flight and evolution power of algorithm with dynamic regenerating population are used to improve WPA algorithm and resolve this problem. Any scales of simulation results are analyzed by experiments and compared with other algorithms. The simulation results show the proposed LDWPA algorithm can acquire near-optimal solutions in reasonable time. Our future work will focus on decreasing algorithm operation time and increasing control requirements for the quality of big scale problems.

**Acknowledgement**

This work was supported by Liaoning Provincial Science Foundation(No. 201602608), Project of Liaoning Province Education Department（No. LJZ2017015）, Shenyang Municipal Science and Technology Project(No. Z18-5-015) and Project of Sichuan Province Education Department (No. 17ZB0823).